\documentclass[prd,twocolumn,showpacs,preprintnumbers,floats,floatfix,
amssymb,amsmath,nofootinbib]{revtex4}
\usepackage{graphicx}
\usepackage{dcolumn}
\usepackage{bm}

\begin{document}
\preprint{CERN-TH/2002-267}

\title{WKB approximation for inflationary cosmological perturbations}

\author{J\'er\^ome Martin} 
\email{jmartin@iap.fr}
\affiliation{Institut d'Astrophysique de Paris, 98bis boulevard Arago,
 75014 Paris, France} 

\author{Dominik J.~Schwarz}
\email{dominik.schwarz@cern.ch}
\affiliation{Theory Division, CERN, 1211 Geneva 23, Switzerland}
\affiliation{Institut f\"ur Theoretische Physik, 
 Technische Universit\"at Wien,
 Wiedner Hauptstra\ss e 8--10, 1040 Vienna, Austria} 

\date{\today} 

\begin{abstract} 
A new method for predicting inflationary cosmological perturbations,
based on the Wentzel--Kramers--Brillouin (WKB) approximation, is
presented. A general expression for the WKB scalar and tensor power
spectra is derived. The main advantage of the new scheme of
approximation is that it is valid even if the slow-roll conditions are
violated. The method is applied to power-law inflation, which allows a
comparison with an exact result. It is demonstrated that the WKB
approximation predicts the spectral indices exactly and the amplitude
with an error lower than $10\%$, even in regimes far from
scale-invariance. The new method of approximation is also applied to a
situation where the slow-roll conditions hold. It is shown that the
result obtained bears close resemblance with the standard slow-roll
calculation. Finally, some possible improvements are briefly
mentioned.
\end{abstract}

\pacs{98.80.Cq, 98.70.Vc}
\maketitle

\section{Introduction}

Observations of the inhomogeneities and anisotropies of the Universe's
large scale structure and of the cosmic microwave background (CMB)
radiation are one of the keys to decipher the cosmic evolution.
According to the favoured cosmological scenario, fluctuations of
matter and space-time are generated during an early epoch of
inflation. The ability to predict the power spectra of cosmological
perturbations for a large variety of different inflationary models is
of the utmost importance. So far, all predictions that come up to the
level of accuracy needed for CMB experiments such as Archeops, MAP or
Planck, have to rely on the slow-roll approximation, subject to an
infinite number of convergence conditions, or on numerical integration
\cite{LLMS}. In this paper we present a new method that overcomes the
mentioned restrictions (but has its own, very different ones).  Our
method is applicable to models that cannot be described by the
slow-roll approximation (e.g. the `problem of a large slow-roll
parameter $\eta$' of some supersymmetry-inspired inflationary models)
and it can be applied to small scales that leave the horizon at times
close to the end of inflation, where the slow-roll approximation
necessarily has to break down.  

\par

The study of scalar and tensor fluctuations can be reduced to the
study of a single variable during inflation, usually denoted by
$\mu_{_{\rm S}}$ or $\mu_{_{\rm T}}$ \cite{GMMFB}. The corresponding
equations of motion are similar to the Schr\"odinger equation for a
one-dimensional system:
\begin{equation}
\label{osci}
\frac{{\rm d}^2}{{\rm d}\eta ^2}\mu +[k^2-U(\eta )]\mu 
=\frac{{\rm d}^2}{{\rm d}\eta ^2}\mu +\omega ^2(\eta )\mu 
=0 \, ,
\end{equation}
where, in the cosmological context, the conformal time $\eta$ plays
the role of the radial variable $r$ in atomic physics. In general,
this equation cannot be solved exactly unless very simple potentials
$U(\eta)$ are considered and one has to rely on a method of
approximation. However, we have inherited a very powerful method to
deal with more complicated potentials from the pioneers of quantum
mechanics, namely the Wentzel--Kramers--Brillouin (WKB)
method of approximation.

\par

In this article, we apply this method to the equation governing
the evolution of cosmological perturbations. It is quite astonishing
that for more than twenty years since the discovery of cosmological
inflation, to our knowledge, nobody has been able to apply
successfully the WKB approximation to the prediction of
cosmological perturbations. This is probably related to the fact
that the growth of the perturbations on superhorizon scales, which is
equivalent, in the quantum-mechanical version of the theory, to
particles creation, is often described in terms of the breakdown of
the WKB approximation. In this article, we demonstrate that this is
true only if the WKB approximation is used naively. In fact, the
problem bears a close resemblance with the situation discussed by
atomic physicists at the time quantum mechanics was
born. We therefore make a historical digression to
illustrate the main point of the present article.

\par

The subject debated by the atomic physicists at the beginning of the
twentieth century was the application of the WKB approximation to the
motion in a central field of force and, more specifically, how the
Balmer formula, for the energy levels of hydrogenic atoms, can be recovered
within the WKB approximation. The effective frequency for hydrogenic
atoms is given by (obviously, in the atomic physics context, the wave
equation is not a differential equation with respect to time but to
the radial coordinate $r$)
\begin{equation}
\label{omeH}
\omega ^2(r)=\frac{2m}{\hbar^2}\biggl(E+\frac{Ze^2}{r}\biggr)
-\frac{\ell (\ell +1)}{r^2}\, ,
\end{equation}
where $Ze$ is the (attractive) central charge and $\ell$ the quantum
number of angular momentum. The symbol $E$ denotes the energy of the
particle and is negative in the case of a bound state. The calculation
of the energy levels was first addressed by Kramers \cite{Kramers} and
by Young and Uhlenbeck \cite{YU}. The method employed was Sommerfeld's
quantization rule
\begin{equation}
\oint \omega (r){\rm d}r= 2 \pi \biggl(n+\frac{1}{2}\biggr) \, ,
\end{equation}
where the integral must be calculated along the classical trajectory
given here by an ellipse. The result obtained by those authors reads
\begin{equation}
\label{Elevel}
E_{n,\ell }=-\frac{mZ^2e^4}{2\hbar ^2}
\biggl\{\biggl(n+\frac{1}{2}\biggr)+[\ell (\ell +1)]^{1/2}
\biggr\}^{-2}\, .
\end{equation}
In addition, the WKB wave function for small values of $r$ was found
to behave as $\simeq r^{[\ell (\ell +1)]^{1/2}+1/2}$. This last result
follows straightforwardly if one retains only the last term in
Eq.~(\ref{omeH}), i.e. $\omega (r)\simeq i[\ell (\ell +1)]^{1/2}/r$,
as required in the limit $r\rightarrow 0$. It was noticed by
Young and Uhlenbeck that the two previous results are not
satisfactory. Indeed, experimental results indicate that the factor
$\ell (\ell +1)$ in Eq.~(\ref{Elevel}) should be replaced by $(\ell
+1/2)^2$: ``{\it It is apparent that we do not obtain the familiar
Balmer formula except in the limit of large $\ell $ values. If,
however, we replace $\ell (\ell +1)$ by $(\ell +1/2)^2$ the correct
result is obtained}.'' A similar remark is made in Ref.~\cite{YU}
concerning the behaviour of the wave function near the origin: ``{\it
The approximate solution based upon the negative exponential behaves
for $r$ very small like $r^{[\ell (\ell +1)]^{1/2}+1/2}$. Again the
replacement of $\ell (\ell +1)$ by $(\ell +1/2)^2$ leads to an
agreement between the approximate and exact solutions}.'' The solution
advocated by Young and Uhlenbeck was therefore to replace, by hand,
$\ell (\ell +1)$ by $(\ell+1/2)^2$ without further theoretical
justifications.  

\par

In $1937$ the problem was considered again by Langer~\cite{Langer}.
In the section devoted to the radial wave equation, he notices that
the WKB approximation was applied to the Schr\"odinger equation
(\ref{osci}) by Kramers, Young and Uhlenbeck only because of its
resemblance with the equation of an oscillator: ``{\it Insofar as the
writer is aware, the degree of this resemblance has hitherto been
regarded without exception as sufficient for assuming the out-and-out
applicability to the equation (17) of the formulas deduced in the
foregoing discussion, especially of the formula (14)},'' where
Eqs.~(17) and (14) in Langer's paper correspond to
Eq.~(\ref{osci}) and the WKB connection formula. He also remarks that
no theoretical explanation has been given to explain why the wrong
factor $\ell (\ell +1)$ is obtained: ``{\it This ``failure'' of the
WKB method, i.e.~to the extent that the change in question is
requisite, has been generally verified both in studies of attractive
and repulsive fields. No explanation of it seems to have been given,
though that can be done very simply to the following effect. The fault
lies not in the method but in the application of it.}'' Then, Langer
shows that the WKB approximation breaks down, at small $r$, for an
effective frequency given by Eq.~(\ref{omeH}) and, in addition, he
suggests a method to circumvent this difficulty. The method consists
in changing the variables according to
\begin{equation}
\label{trans}
r={\rm e}^x, \quad \mu ={\rm e}^{x/2}u \,.
\end{equation}
The resulting equation for $u$ has again the form of Eq.~(\ref{osci})
but the effective frequency is now given by
\begin{equation}
\omega^2(x)=\frac{2m}{\hbar ^2}(E {\rm e}^{2x}+Ze^2{\rm e}^x)
-\biggl(\ell +\frac{1}{2}\biggr)^2 \, .
\end{equation}
As demonstrated by Langer, the crucial point is that, now, the WKB
approximation, applied to the Schr\"odinger equation for $u(x)$, no
longer breaks down. Therefore, it is now possible to utilize the
WKB formalism and to use the usual WKB approximation. Transforming
back to the original variables the correct factor $(\ell+1/2)^2$
appears, instead of the wrong term $[\ell (\ell +1)]^{1/2}$. This
allows Langer to conclude that his method ``{\it ... is seen to amount
formally to precisely the replacement of $\ell (\ell +1)$ by $(\ell
+1/2)^2$}''.

\par

Let us now come back to the theory of cosmological perturbations. The
analogy with the previous situation is that large values of $r$
correspond to subhorizon scales ($k |\eta| \gg 1$) whereas small
values of $r$ correspond to superhorizon scales ($k|\eta| \ll 1$).  On
subhorizon scales the mode function oscillates ($\omega \simeq k$) and
the WKB approximation can be applied in this regime without any
difficulty. On superhorizon scales the perturbations do not oscillate,
but this is not a problem in itself. The WKB approximation can be used
in a situation where the potential energy dominates and leads to
exponentially decaying/growing solutions (e.g. in the WKB treatment
of the $\alpha$ decay of a radioactive nucleus).  Nevertheless,
difficulties arise if one tries to find the WKB approximation to the
mode equations of cosmological perturbations.

\par 

On superhorizon scales a problem arises in the naive WKB approach
{\it \`a la} Kramers, Uhlenbeck and Young from the form of the
effective potential $U(\eta )$. It turns out to be the very same
problem as encountered by atomic physicists. The resemblance is
most explicit in the particular case of power-law inflation where the
potential $U(\eta )\propto \eta ^{-2}$ is exactly of the form
given in Eq.~(\ref{omeH}). The usual conclusion that the WKB
approximation is violated for superhorizon scales is in fact not true,
provided the cosmological version of the transformation~(\ref{trans})
can be found. This opens the possibility to employ the WKB
approximation to calculate the spectrum of inflationary cosmological
perturbations in the superhorizon limit. In the cosmological context,
the WKB approximation is a new method for computing the power spectra,
different from the more traditional slow-roll approximation.

\par

This article is organized as follows. In the next section, we briefly
recall the basic results of the inflationary cosmological
perturbations theory. In the third section, we show how the WKB
approximation can be applied to inflationary cosmological
perturbations in the general case, i.e.~for any inflaton
potential. The only restriction is that we focus on single inflaton
field models. We demonstrate that the calculation of the spectrum
reduces to the calculation of a single quadrature. In the fourth
section we apply the WKB approximation to power-law inflation. This
model is exactly solvable so that we can compare the exact result with
the WKB prediction and also with the standard calculation based on the
slow-roll approximation. This allows us to study the precision of the
WKB approximation in the cosmological context. In particular, we
demonstrate that, in the regime where the slow-roll approximation
breaks down (far from scale invariance), the WKB approximation is
still valid. This opens the possibility to calculate the power spectra
of more realistic models violating the slow-roll conditions. In the
fifth section, we apply the WKB approximation to the general 
(first-order) slow-roll situation. We compare the result with the standard
slow-roll calculation for the amplitudes and the spectral indices. We
show that the spectral indices are very well predicted while the error
in the amplitude is of the order $10\%$, i.e. less accurate than
the prediction of the slow-roll approximation for this
quantity. Finally, in the sixth section, we give our conclusions and
indicate what the possible extensions to the present work are.

\section{General expressions for inflationary perturbations}

\subsection{Power spectra}

The evolution of the cosmological perturbations in a single field
inflationary scenario is governed by the Schr\"odinger-like 
equation~(\ref{osci}). The effective time-dependent frequency is given
by the general expression 
\begin{equation}
\label{eqmu}
\omega ^2(k,\eta ) =  k^2-\frac{z''}{z},
\end{equation}
where the primes denote a derivatives with respect to conformal time and
where $z_{_{\rm S}}= a\sqrt{2-aa''/a'^2}$ for scalar (density)
perturbations and $z_{_{\rm T}}=a$ for tensor perturbations; 
$a(\eta )$ is the scale factor of the
Friedmann--Lemaitre--Robertson--Walker metric. For density
perturbations, the quantity $\mu$ is given by $\mu_{_{\rm S}}= - 2
z_{_{\rm S}} \zeta$, where $\zeta$ is Bardeen's
hypersurface-independent quantity \cite{Bardeen}, which for adiabatic
perturbations is conserved on superhorizon scales.  For tensor
fluctuations, one simply has $\mu_{_{\rm T}}=z_{_{\rm T}}h$, where $h$
is the amplitude of gravitational waves. The equation of motion for
the mode functions $\mu$ should be solved with the following initial
condition
\begin{equation}
\label{ini}
\lim _{k/(aH)\rightarrow +\infty}\mu _{_{\rm S,T}}(\eta )=\mp
\frac{4\sqrt{\pi }}{m_{_{\rm Pl}}}
\frac{{\rm e}^{-ik(\eta -\eta _{\rm i})}}{\sqrt{2k}},
\end{equation}
where $\eta_{\rm i}$ is an arbitrary time at the beginning of
inflation and $m_{_{\rm Pl}}$ is the Planck mass. This initial
condition corresponds to the fact that, initially, the modes are
subhorizon and, therefore, do not feel the curvature of space-time. As
a consequence, they are described by plane waves. If the initial
quantum state is the vacuum state, then the statistical properties of
the perturbations are entirely characterized by the two-point
correlation function, i.e.~by the power spectrum. The dimensionless
power spectra of scalar and tensor fluctuations are calculated to read
\begin{equation}
\label{pspec}
{\cal P}_\zeta = \frac{k^3}{8\pi ^2}\biggl\vert 
\frac{\mu _{_{\rm S}}}{z_{_{\rm S}}}
\biggr \vert ^2,\quad 
{\cal P}_h = \frac{2k^3}{\pi ^2}\biggl\vert 
\frac{\mu _{_{\rm T}}}{z_{_{\rm T}}}
\biggr \vert ^2.
\end{equation}
The spectral indices and their running are defined by the coefficients
of Taylor expansions of the power spectra with respect to $\ln k$,
evaluated at an arbitrary pivot scale $k_*$. To be more specific: 
\begin{equation}
\label{n}
n_{_{\rm S}} -1 \equiv {{\rm d} \ln {\cal P}_\zeta\over {\rm d} \ln k}
\biggr\vert_{k=k_*},
\quad
n_{_{\rm T}} \equiv {{\rm d} \ln {\cal P}_h\over {\rm d} \ln k}
\biggr\vert_{k=k_*} \, ,
\end{equation}
are the spectral indices and the two following expressions
\begin{equation}
\label{alpha}
\alpha_{_{\rm S}} \equiv {{\rm d}^2 \ln {\cal P}_\zeta\over 
({\rm d} \ln k)^2}\biggr\vert_{k=k_*}, 
\quad
\alpha_{_{\rm T}} \equiv  {{\rm d}^2 \ln {\cal P}_h\over ({\rm d} \ln k)^2}
\biggr\vert_{k=k_*}\, , 
\end{equation}
define the ``running'' of these indices. In principle, we could
also define the running of the running and so on.

\subsection{Inflationary parameters}

In order to investigate the predictions of cosmological inflation in a
model-independent manner, it is useful to define a set of functions
that uniquely characterize the state of the Universe at any given
moment of time. It is usefull to describe the evolution of the Universe
by a set of flow equations \cite{HT,STG}. In this article we make use of 
the set of horizon flow functions, introduced in Ref.~\cite{STG}. 
The zeroth horizon flow
function is defined by $\epsilon_0\equiv H(N_{\rm i})/H(N)$, where
$H\equiv \dot{a}/a$ is the Hubble rate, a dot meaning derivative with
respect to cosmic time. In this expression, $N$ is the number of
e-folds, $N\equiv \ln (a/a_{\rm i})$, after an arbitrary initial
time. The hierarchy of horizon flow functions is then defined
according to
\begin{equation}
\label{flow}
\epsilon _{n+1}\equiv 
\frac{{\rm d}\ln \vert \epsilon _n\vert }{{\rm d}N}, \quad 
n\ge 0\, .
\end{equation}
Inflation takes place for $\epsilon_1 < 1$.  
It is interesting to establish the link between the horizon flow
functions and the so-called slow-roll parameters, the first three
being defined by
\begin{eqnarray}
\label{defsrpara}
\epsilon & \equiv & 3\frac{\dot{\varphi }^2}{2}\biggl[
\frac{\dot{\varphi }^2}{2}+V(\varphi )\biggr]^{-1}
=-\frac{\dot{H}}{H^2},
\\
\delta &\equiv & -\frac{\ddot{\varphi }}{H\dot{\varphi }}=
-\frac{\dot{\epsilon }}{2H\epsilon }+\epsilon, 
\\ 
\xi &\equiv & \frac{\dot{\epsilon}-\dot{\delta }}{H},
\end{eqnarray}
where $\varphi $ is the inflaton field and $V(\varphi )$ is the
inflaton potential. The first two slow-roll parameters must be small
during inflation, as can be seen from their definition. The parameter
$\epsilon$ is just the inflaton kinetic energy to inflaton total
energy ratio, whereas $\delta$ is the acceleration to speed ratio. The
set $\{\epsilon , \delta , \xi \}$ is linked to the horizon flow
functions $\{\epsilon _n\}$ by
\begin{equation}
\epsilon =\epsilon _1\, ,\quad 
\delta =\epsilon _1-\frac{1}{2}\epsilon _2\, ,
\quad
\xi =\frac{1}{2}\epsilon _2 \epsilon _3 \, .
\end{equation}

The set $\{\epsilon , \delta , \xi \}$ or equivalently the set
$\{\epsilon_1, \epsilon_2, \epsilon_3\}$ plays a special role, since
the effective potential $U(\eta )\equiv z''/z$ of Eq.~(\ref{osci}) can
be expressed exactly in terms of these three functions only. One finds
\begin{eqnarray}
\label{potsr}
U_{_{\rm S}}\! &=& \! a^2\! H^2 (2 - \epsilon_1 +
{\textstyle\frac 32}\epsilon_2 - {\textstyle\frac 12}\epsilon_1\epsilon_2 +
{\textstyle\frac 14}\epsilon_2^2 + {\textstyle\frac 12}\epsilon_2\epsilon_3),
\\ 
U_{_{\rm T}}\! &=& \! a^2\! H^2(2-\epsilon_1).
\end{eqnarray}
Let us emphasize that these expressions are exact, no approximation
has been used so far. \par

\section{WKB power spectra}

We now turn to the derivation of the WKB inflationary power spectra
for scalar and tensor perturbations. We define the quantity
$\mu_{_{\rm WKB}}$ by the expression
\begin{equation}
\label{wkbsol}
\mu_{_{\rm WKB}}(k,\eta )\equiv \frac{1}{\omega^{1/2}(k,\eta)}
{\rm e}^{\pm i\int^{\eta }\omega(k,\tau){\rm d}\tau}\, .
\end{equation}
The mode function $\mu_{_{\rm WKB}}$ represents the leading order term of
the semiclassical expansion. It satisfies the following differential equation
\begin{equation}
\label{eqwkb}
\mu _{_{\rm WKB}}''(k,\eta ) +\biggl[\omega ^2(k,\eta )-
Q(k,\eta )\biggr]\mu_{_{\rm WKB}}(k,\eta )=0\, ,
\end{equation}
where the quantity $Q(k,\eta )$ is given by
\begin{equation}
Q(k,\eta )\equiv 
\frac{3}{4}\frac{(\omega ')^2}{\omega ^2}
-\frac{\omega ''}{2\omega }\, .
\end{equation}
Therefore, the mode function $\mu _{_{\rm WKB}}(k,\eta )$ given in
Eq.~(\ref{wkbsol}) is a good approximation of the actual mode function
$\mu (k,\eta )$, if the following condition is satisfied
\begin{equation}
\biggl \vert \frac{Q}{\omega^2}\biggr \vert \ll 1\, .
\end{equation}
On subhorizon scales, $\omega \simeq k$, which implies $Q\simeq 0$ and
therefore the condition is satisfied. On superhorizon scales, it is
not possible to show that the previous condition is always violated,
since it depends on the shape of the effective potential $U(\eta)$.
However, in the cosmological context, the problem discussed in the 
introduction generally arises. We explicitly show this in the
next section for the case of power-law inflation. Here, we demonstrate
the problem for the more generic case of slow-roll inflation. Using
the expression for the effective frequency in terms of the horizon
flow functions, one has
\begin{equation}
\biggl \vert \frac{Q}{\omega^2}\biggr \vert _{_{\rm S}}
=\biggl \vert \frac{Q}{\omega^2}\biggr \vert _{_{\rm T}}
=\frac{1}{8}+{\cal O}(\epsilon_1,\epsilon_2)\, .
\end{equation}
The corrections, linear in the horizon flow functions, could easily
multiply the term $1/8=0.125$ by a factor $2$. Therefore, the number
$\vert Q/\omega ^2\vert $ is indeed not greater than $1$ but, at the
same time, not very small with respect to $1$. Thus, the WKB
approximation is not a good approximation in this regime. In order to
have an effective potential that permits to use the WKB
approximation, we introduce the following transformation
\begin{equation}
\label{transf}
x\equiv \ln \biggl(\frac{Ha}{k}\biggr), \quad 
u\equiv (1-\epsilon_1)^{1/2}{\rm e}^{x/2}\mu .
\end{equation}
This is the cosmological counterpart of Langer's
transformation given in the introduction; see
Eq.~(\ref{trans}). However, let us remark that there exist
differences, in particular the factor $(1-\epsilon_1)^{1/2}$ in the
definition of the function $u$. It is also interesting to notice
that this transformation has already been considered in the literature
in Ref.~\cite{WMS} for different reasons. Then, for density
perturbations, the equation of motion takes the form
\begin{widetext}
\begin{equation}
\label{eqmox}
\frac{{\rm d}^2u_{_{\rm S}}(x)}{{\rm d}x^2}+
\biggl[\frac{{\rm e}^{-2x}}{(1-\epsilon_1)^2}
-\frac{1}{4}\biggl(\frac{3-\epsilon_1}{1-\epsilon_1}\biggr)^2
-\frac{(3-2\epsilon_1)\epsilon_2}{2 (1-\epsilon_1)^2}
-\frac{(1-2\epsilon_1)\epsilon_2\epsilon_3}{2(1-\epsilon_1)^3}
-\frac{(1-4\epsilon_1)\epsilon_2^2}{4(1-\epsilon_1)^4}
\biggr]u_{_{\rm S}}(x)=0,
\end{equation}
whereas, for gravitational waves, we obtain
\begin{equation}
\label{eqmoxgw}
\frac{{\rm d}^2u_{_{\rm T}}(x)}{{\rm d}x^2}+
\biggl[\frac{{\rm e}^{-2x}}{(1-\epsilon_1)^2}
-\frac{1}{4}\biggl(\frac{3-\epsilon_1}{1-\epsilon_1}\biggr)^2
+\frac{\epsilon_1\epsilon_2}{2(1-\epsilon_1)^2}
+\frac{\epsilon_1\epsilon_2\epsilon_3}{2(1-\epsilon_1)^3}
+\frac{(2+\epsilon_1)\epsilon_1\epsilon_2^2}{4(1-\epsilon_1)^4}
\biggr]u_{_{\rm T}}(x)=0.
\end{equation}
\end{widetext}
In the following $\omega_{_{\rm S}}^2$ and $\omega_{_{\rm T}}^2$
denote the expressions in the square brackets of Eqs. (\ref{eqmox})
and (\ref{eqmoxgw}). The WKB approximation can now be applied to
these equations. In order to demonstrate this, let us recalculate, 
on superhorizon scales, the quantity $\vert Q/\omega ^2\vert $
for the two previous expressions. One finds for density
perturbations
\begin{equation}
\label{testdp}
\biggl \vert \frac{Q}{\omega^2}\biggr \vert _{_{\rm S}}
=\frac{4}{27}\biggl(\epsilon_1\epsilon_2^2 + \epsilon_1\epsilon_2\epsilon_3 +
\frac{\epsilon_2\epsilon_3^2}{2} + \frac{\epsilon_2\epsilon_3\epsilon_4}{2}
\biggr)+{\cal O}(\epsilon_n^4)\, ,
\end{equation}
whereas for gravitational waves, one has
\begin{equation}
\label{testgw}
\biggl \vert \frac{Q}{\omega^2}\biggr \vert _{_{\rm T}}
=\frac{4}{27}(\epsilon_1\epsilon_2^2 + \epsilon_1\epsilon_2\epsilon_3)+
{\cal O}(\epsilon_n^4)\, .
\end{equation}
It is obvious that the quantity $\vert Q/\omega ^2 \vert $ is now very
small for any inflaton potential satisfying the slow-roll conditions,
since this is a third-order quantity. As a consequence, the WKB
approximation is now also valid on superhorizon scales.

\par

Let us apply the WKB approximation to Eqs.~(\ref{eqmox}) and (\ref{eqmoxgw}).
Let $x_*$ be the value of the variable $x$ such that $\omega ^2(x_*)=0$, 
i.e. $x_*$ is the classical turning point. We define region I as the region 
such that $\omega^2(x)>0$ and region II the region where $\omega ^2(x)<0$. 
Let $x_{\rm i}$ be the initial ``time'' at which the normalization is 
performed (i.e. corresponding to $\eta _{\rm i}$). At some point $x$, deep in
region I, the WKB approximation of the function $u(x)$ can be written as 
\begin{equation}
\label{wkbuI}
u_{_{\rm I}}(x)\simeq \frac{A}{\omega ^{1/2}(x)}\exp
\biggl[\pm i\int _{x_{\rm i}}^{x}\omega (y){\rm d}y\biggr]\, ,
\end{equation}
where $A$ is a constant of normalization. Deep in region I, we have
$\omega (y)\simeq e^{-y}/(1-\epsilon_1)$ and straightforward
manipulations show that the corresponding $\mu_{_{\rm I}}(\eta )$
behaves as in Eq. (\ref{ini}), provided that one chooses the minus
sign in Eq. (\ref{wkbuI}) and that $A$ is given by
\begin{equation}
\label{A}
A=\mp\frac{4}{m_{_{\rm Pl}}}\sqrt{\frac{\pi }{2}}k^{-1/2}.
\end{equation}
In region II, the effective frequency is pure imaginary and can be
written as $\omega (x)=i|\omega (x)|$. Let $x_{\rm f}$ be the final
point at which we evaluate the solution deep in region II. Then, at
any point $x$ in region II, the WKB solution is given by
\begin{eqnarray}
\label{wkbuII}
u_{_{\rm II}}(x) &\simeq& \frac{C_+}{|\omega (x)|^{\frac 12}}
  \exp\biggl[+\int _{x}^{x_{\rm f}}\! |\omega(y)|{\rm d}y\biggr] 
  \nonumber \\
& & + \frac{C_-}{|\omega (x)|^{\frac 12}}
  \exp\biggl[-\int_{x}^{x_{\rm f}}\! |\omega(y)|{\rm d}y\biggr] ,
\end{eqnarray}
where $C_{\pm}$ are constant. The goal is now to connect these two
constants with the constant $A$. At the turning point $x=x_*$, the WKB
approximation breaks down and one has to use the usual WKB procedure
and approximate the potential in this region by a straight line such
that
\begin{equation}
\label{linearw}
\omega ^2(x)\simeq -\alpha (x-x_*),
\end{equation}
where $\alpha \equiv -[{\rm d}\omega ^2(x)/{\rm d}x](x=x_*)>0$. The
solutions of Eqs. (\ref{eqmox}) and (\ref{eqmoxgw}) are given in terms
of Airy functions of first and second kinds
\begin{equation}
\label{sollin}
u(x)=B_1 {\rm Ai}(s)+B_2 {\rm Bi}(s),
\end{equation}
where $s\equiv \alpha ^{1/3}(x-x_*)$. 

\begin{widetext}
The next move is to use the asymptotic behaviour of the Airy functions
to calculate the relation between $B_1$ and $B_2$ with $A$ on one hand
and $C_{\pm}$ on the other. In region I, for a value of $x$ not too
far from $x_*$, Eq. (\ref{sollin}) can be written as
\begin{equation}
u_{_{\rm I}}(x) \simeq \frac{\alpha^{-\frac{1}{12}}}{2\sqrt{\pi}}
|x-x_*|^{-\frac 14}
\biggl[(B_2-iB_1)\exp 
\biggl(\frac{i2}{3}\alpha^{\frac 12}|x-x_*|^{\frac 32}+\frac{i\pi}{4}\biggr)
+ (B_2+iB_1)\exp 
\biggl(-\frac{i2}{3}\alpha^{\frac 12}|x-x_*|^{\frac 32}-\frac{i\pi}{4}\biggr)
\biggr],
\end{equation}
whereas in region II, under the same conditions, the function
$u(x)$ can be expressed as
\begin{equation}
\label{AiII}
u_{_{\rm II}}(x)\simeq \frac{\alpha ^{- \frac{1}{12}}}{\sqrt{\pi }}
(x-x_*)^{-\frac 14}
\biggl\{\frac{B_1}{2}\exp \biggl[-\frac{2}{3}
\alpha^{\frac 12}(x-x_*)^{\frac 32}\biggr]
+B_2\exp \biggl[\frac{2}{3}\alpha^{\frac 12}(x-x_*)^{\frac 32}\biggr]
\biggr\}.
\end{equation}
\end{widetext}
There is no factor $1/2$ in front of the coefficient $B_2$ because
the expansion of the function ${\rm Bi}$ differs from the expansion of
${\rm Ai}$ not only in the sign of the argument of the exponential but
also by a factor $1/2$. Now we evaluate in region I, the approximate
solution (\ref{wkbuI}) for the potential (\ref{linearw}).  The
integral in the exponent can be written as
\begin{equation}
\label{int}
\int_{x_{\rm i}}^{x}\!\omega(y){\rm d}y=
\int_{x_{\rm i}}^{x_*}\!\omega (y){\rm d}y
+\int_{x_*}^{x}\!\omega (y){\rm d}y
\equiv \Phi +\int_{x_*}^x\!\omega (y){\rm d}y.
\end{equation}
The frequency (\ref{linearw}) is used in the second integral
only, assuming that $x$ is not too far away from $x_*$. The quantity
$\Phi$ is just a number and its calculation would require the
knowledge of $\omega(x)$ in the whole region I. However, it does not
enter the final result and therefore we are not interested in its
value. We find
\begin{equation}
\label{wkbI*}
u_{_{\rm I}}(x)=A\alpha^{-\frac 14}|x-x_*|^{- \frac 14}
\exp \biggl(\frac{i2}{3}\alpha^{\frac 12}|x-x_*|^{\frac 32} + i\Phi\biggr)\, ,
\end{equation}
from which we deduce 
\begin{equation}
\label{linkAB}
B_1=iB_2, \quad B_2=A\sqrt{\pi}\alpha^{-1/6}{\rm e}^{i(\Phi - \pi /4)}.
\end{equation}
Therefore, we have reached our first goal, i.e. connect the two
coefficients $B_1$ and $B_2$ to $A$. What remains to be done is to
establish the link between $C_+$, $C_-$ and $B_1$, $B_2$. This can be
done by applying the very same reasoning in region II. One introduces
$\Psi$, defined by
\begin{equation}
\label{defpsi}
\Psi \equiv \int _{x_*}^{x_{\rm f}}|\omega (y)|{\rm d}y\, ,
\end{equation}
and the WKB solution given in Eq.~(\ref{wkbuII}) reduces to the expression
\begin{widetext}
\begin{equation}
\label{wkbII*}
u_{_{\rm II}}(x)=\alpha^{-\frac 14}(x-x_*)^{- \frac 14}\biggl\{C_+e^{\Psi}
\exp \biggl[-\frac{2}{3}\alpha ^{\frac 12}(x-x_*)^{\frac 32}\biggr]
+C_-e^{-\Psi }
\exp \biggl[\frac{2}{3}\alpha^{\frac 12}(x-x_*)^{\frac 32}\biggr]\biggr\}.
\end{equation}
\end{widetext}
By comparison with Eq.~(\ref{AiII}), the expressions for $C_+$ and
$C_-$ are obtained
\begin{equation}
\label{linkBC}
C_+=\frac{B_2}{2\sqrt{\pi }}\alpha ^{1/6}{\rm e}^{-\Psi +i\pi /2}, \quad 
C_-=\frac{B_2}{\sqrt{\pi }}\alpha ^{1/6}{\rm e}^{\Psi }.
\end{equation}
Finally, putting everything together, we arrive at
\begin{equation}
\label{Cfinal}
|C_+|^2=\frac{2\pi }{m_{_{\rm Pl}}^2}{\rm e}^{-2\Psi}k^{-1}, \quad 
|C_-|^2=\frac{8\pi }{m_{_{\rm Pl}}^2}{\rm e}^{2\Psi}k^{-1}.
\end{equation}
If we only consider the growing mode given by the branch proportional
to $C_-$, which is a good approximation for $x \gg x_*$, one can
deduce the power spectra at $x=x_{\rm f}$. Using Eqs.~(\ref{pspec}),
one obtains the dimensionless power spectra within the WKB
approximation:
\begin{eqnarray}
\label{wkbpspecdp}
{\cal P}_{\zeta } &=& 
\frac{H^2}{\pi \epsilon _1m_{_{\rm Pl}}^2}\biggl(\frac{k}{aH}\biggr)^3
\frac{{\rm e}^{2\Psi _{_{\rm S}}}}{(1-\epsilon _1)|\omega _{_{\rm S}}|}, 
\\
\label{wkbpspecgw}
{\cal P}_{h} &=& \frac{16H^2}{\pi m_{_{\rm Pl}}^2}
\biggl(\frac{k}{aH}\biggr)^3
\frac{{\rm e}^{2\Psi _{_{\rm T}}}}{(1-\epsilon _1)|\omega _{_{\rm T}}|},
\end{eqnarray}
where all the quantities in the above expression are evaluated at
$x=x_{\rm f}$. We know that in the absence of entropy perturbations and 
decaying modes, $\zeta$ is conserved on superhorizon scales \cite{Bardeen,MS}. 
Similarly, $h$ is conserved on superhorizon scales, if the decaying mode 
is absent. Thus we know that both spectra should be time-independent. However,
within the WKB approximation one cannot see this from the above expressions. 
In the following sections we demonstrate this property explicitly for 
power-law inflation and slow-roll inflation. 

We see that the calculation of the spectrum for any
inflaton potential has been reduced to the calculation of a
quadrature, namely $\Psi$. We are also in a position to calculate
the spectral indices and their running. The spectral indices read
\begin{equation}
n_{_{\rm S}} -1 = 3 + 2 {{\rm d} \Psi_{_{\rm S}}\over {\rm d}\ln k}
\biggr\vert_{k=k_*}\, ,
\quad
n_{_{\rm T}} = 3 + 2 {{\rm d} \Psi_{_{\rm T}}\over {\rm d}\ln k}
\biggr\vert_{k=k_*}\, , 
\end{equation}
and the runnings are given by 
\begin{equation}
\alpha_{_{\rm S}} = 2 {{\rm d}^2 \Psi_{_{\rm S}}\over ({\rm d}\ln k)^2} 
\biggr\vert_{k=k_*},
\quad
\alpha_{_{\rm T}} = 2 {{\rm d}^2 \Psi_{_{\rm T}}\over ({\rm d}\ln k)^2} 
\biggr\vert_{k=k_*}\, .
\end{equation}
The general expressions for the WKB power spectra, spectral indices
and their runnings, constitute the main result of this work.  In order
to discuss the power and potential of the WKB approximation, we now
turn to two cases where the quadratures $\Psi_{_{\rm S}}$ and
$\Psi_{_{\rm T}}$ can be calculated explicitly: power-law and
slow-roll inflation.

\section{Application to power-law inflation}

We first apply the formalism developed in the previous section to 
power-law inflation. In this model, the scale factor is given by
\begin{equation}
\label{scalef}
a(\eta )=\ell_0|\eta |^{1+\beta },
\end{equation}
where $\beta \le -2$. This scale factor is a solution of the Einstein
equations in the case where the inflaton potential is $V \propto
\exp(\varphi /m_{_{\rm Pl}})$. The case $\beta =-2$ is special, since
it corresponds to the de Sitter space-time. The Hubble radius is then
constant and given by $\ell_0$. In the general case, the horizon flow 
functions parameters are easily calculated and read
\begin{equation}
\label{srpl}
\epsilon_1 = \frac{2+\beta }{1+\beta }, \quad \epsilon_n = 0,  
\quad n>1\, .
\end{equation}
The exact power spectrum can be found and is given by \cite{AW,MS}
\begin{eqnarray}
\label{pspldp}
{\cal P}_{\zeta} &=& \frac{\ell_{_{\rm Pl}}^2}{\ell_0^2}
\frac{1}{\pi \epsilon _1}f(\beta )k^{2\beta +4}\, ,
\\
\label{psplgw}
{\cal P}_h &=& \frac{\ell_{_{\rm Pl}}^2}{\ell_0^2}\frac{16}{\pi }
f(\beta )k^{2\beta +4}\, ,
\end{eqnarray}
where $\ell _{_{\rm Pl}}=m_{_{\rm Pl}}^{-1}$ is the Planck length
and the function $f(\beta )$ is given by
\begin{equation}
\label{f}
f(\beta )\equiv \frac{1}{\pi }
\biggl[\frac{\Gamma (-\beta -1/2)}{2^{\beta +1}}\biggr]^2.
\end{equation}
In the previous expression, $\Gamma$ denotes Euler's integral of the
second kind. We have $f(\beta=-2)=1$. The case $\beta =-2$ is singular,
since $\epsilon _1=0$ and the expression of the scalar power
spectrum blows up. This case should be considered separately and one
can show that there are no density perturbations at all in the de
Sitter space-time~\cite{MS}. The spectral indices and runnings can
also be calculated very easily. One finds $n_{_{\rm S}}-1=n_{_{\rm
T}}=2\beta +4$ and $\alpha _{_{\rm S}}=\alpha _{_{\rm T}}=0$. Exact
scale-invariance is obtained for $\beta =-2$, which is precisely the
singular case evoked above.  

\par

Let us now apply the WKB approximation to the case under
consideration. First, suppose that we want to use it before doing the
transformation (\ref{transf}).  In this case, we have $\omega ^2(\eta
)=k^2-\beta (\beta +1)/\eta ^2$. In the limit where $k\eta $ goes to
zero, it is easy to see that
\begin{equation}
\label{testwkbpl}
\left\vert\frac{Q}{\omega ^2}\right\vert =\frac{1}{4|\beta (\beta +1)|}\, .
\end{equation}
In general the condition $|Q/\omega^2|\ll 1$ is not satisfied unless
$- \beta$ goes to infinity, an unrealistic situation. For $\beta =-2$,
one recovers that $\vert Q/\omega ^2\vert =1/8$. This is in agreement
with the calculation of $\vert Q/\omega ^2\vert $ performed previously
in the slow-roll approximation since this approximation is,
very roughly speaking, an expansion around the scale-invariant
solution. Equation (\ref{testwkbpl}) shows that the WKB approximation
cannot be applied to the usual equation of motion without first performing
the transformation (\ref{transf}). This conclusion is reinforced
by the following argument. The solution on superhorizon scales is
exactly known and can be expressed as
\begin{equation}
\label{supersol}
\mu _{_{\rm II}}(\eta )=C_1z(\eta )+
C_2 z(\eta )\int ^{\eta }\frac{{\rm d}\tau}{z^2(\tau)},
\end{equation} 
where $C_1$ and $C_2$ are two arbitrary constants. The first branch is
the growing mode whereas the second one is the decaying mode. For
power-law inflation this gives a mode proportional to $\mu \propto
|\eta |^{1+\beta }$ and one proportional to $\mu \propto |\eta
|^{-\beta }$. One can wonder what the link between this solution and
the WKB solution is in region II. On superhorizon scales, we have
$|\omega |\simeq [\beta (\beta +1)]^{1/2}/\vert \eta \vert$. As a
consequence, the mode function $\mu _{_{\rm WKB}}$ has the form $\mu
_{_{\rm WKB}}\propto \vert\eta \vert^{1/2\pm [\beta(\beta
+1)]^{1/2}}$. This is not the correct behaviour given by the previous
exact solution. The correct solution is obtained by replacing the
factor $\beta (\beta +1)$ by $(\beta +1/2)^2$, i.e.~exactly the remark
made by Kramers, Young and Uhlenbeck (see the Introduction). The
reason is very clear: the effective power-law potential has exactly
the same shape as the one describing the motion in a central-force
field, provided that the identification $\ell \leftrightarrow \beta $
is made. Let us also remark that the form of $\mu_{_{\rm WKB}}$ is the
same as the wave function for small $r$, as expected from the previous
considerations. This confirms that the WKB approximation cannot be
used naively on superhorizon scales.

\par

The cure is to use the transformation given in Eq.~(\ref{transf}). After 
this, the effective frequency takes the form
\begin{equation}
\label{omegapl}
\omega _{_{\rm S}}^2(x)=\omega _{_{\rm T}}^2(x)=
(1+\beta )^2{\rm e}^{-2x}-\biggl(\beta +\frac{1}{2}\biggr)^2.
\end{equation}
In the superhorizon limit, $x \to \infty$, one now has $Q \to 0$,
indicating that the WKB solution gives a good approximation. In
addition, the mode function now has the correct time
dependence. Indeed, since $\int _{x}^{x_{\rm f}}|\omega |{\rm d}y=
|\beta +1/2|(x_{\rm f}-x)$, returning to the dependence in conformal
time gives $u_{_{\rm II}} \propto |\eta |^{\pm (\beta +1/2)}$.
Taking into account the link between $u$ and $\mu $, one finally
arrives at $\mu _{_{\rm II}}\propto |\eta |^{1+\beta }$ or 
$\mu _{_{\rm II}}\propto |\eta
|^{-\beta }$, i.e. the growing and decaying modes of
Eq.~(\ref{supersol}) in the case of power-law inflation.  This means
that, in this case, the superhorizon solution is nothing but the WKB
solution.  However, this need not be true for more general
models. A change of variables, which allows us to come back from $x$
to the conformal time $\eta $, shows that the quantity $\Psi $ is
given by
\begin{equation}
\label{psipl}
\Psi =-\int _{\rm k\eta _{\rm f}}^{k\eta _*}\frac{{\rm d}(k\eta )}{k\eta }
\sqrt{-k^2\eta ^2+\biggl(\beta +\frac{1}{2}\biggr)^2},
\end{equation}
where $\eta _*$ can now be expressed as $k\eta _*=\beta +1/2$. This
integral can be performed exactly, for instance with the help of
formula (2.275.3) of Ref. \cite{GR}. One obtains
\begin{widetext}
\begin{eqnarray}
\label{intpl}
\Psi &=& -\biggl \{\sqrt{-k^2\eta _{\rm f}^2+
\biggl(\beta +\frac{1}{2}\biggr)^2}
+\frac{1}{2}\biggl \vert \beta +\frac{1}{2}\biggr \vert
\ln \biggl[\frac{|\beta +1/2|-\sqrt{-k^2\eta _{\rm f}^2+(\beta +1/2)^2}}
{|\beta +1/2|+\sqrt{-k^2\eta _{\rm f}^2+(\beta +1/2)^2}}\biggr]\biggr \} \\
&\simeq & -\biggl \vert \beta +\frac{1}{2}\biggr \vert \biggl [
1+\ln \biggl(\frac{|k\eta _{\rm f}|}{2|\beta +1/2|}\biggr )\biggr ],
\end{eqnarray}
where in the last expression we have considered that $|k\eta _{\rm
f}|\ll 1$. Inserting this result in Eqs.~(\ref{wkbpspecdp}) and
(\ref{wkbpspecgw}), one finds
\begin{equation}
\label{pspecplwkb1}
{\cal P}_{\zeta }=\frac{\ell_{\rm Pl}^2}{\ell_0^2}
\frac{1}{\pi \epsilon_1}g(\beta)k^{2\beta +4},\quad 
{\cal P}_{h}=\frac{\ell_{\rm Pl}^2}{\ell_0^2}\frac{16}{\pi }
g(\beta)k^{2\beta +4}, \quad
g(\beta) \equiv \frac{2{\rm e}^{2\beta +1}}{(2\beta +1)^{2\beta +2}}\, .
\end{equation}
These expressions are the main result of this section. They should
be compared to the exact result given in Eqs.~(\ref{pspldp}) and
(\ref{psplgw}). As expected, they are time-independent.
\end{widetext}
The first observation is that the shape of the power spectra is exactly 
reproduced by the WKB approximation. In particular, this means that the 
spectral indices and their runnings (as well as the runnings of the runnings,
and so on) are predicted exactly by the WKB approximation. The
amplitude is given by the function $g(\beta )$, whereas in the exact
case, it is given by the function $f(\beta )$; see Eq.~(\ref{f}). We
have $g(\beta =-2)=18\times {\rm e}^{-3}\simeq 0.896$, a value that
should be compared with $f(\beta =-2)=1$. In the limit where $|\beta
|$ goes to infinity we can use the Stirling formula, $\Gamma (x)\simeq
\sqrt{2\pi } x^{x-1/2}{\rm e}^{-x}$ to approximate $f(\beta )$ and one
finds that $\lim_{\beta \to - \infty} f(\beta )\to g(\beta )$,
i.e.~the WKB approximation gives the exact result in this limit. It is
interesting to notice that this limit corresponds to going away from
scale invariance, i.e.~to ``fast-roll'' inflationary models (the
kinetic energy of the inflaton is of the same order as its potential
energy). Therefore, the power-law inflation with $\vert \beta \vert \gg 2$
is an explicit example of a model where the WKB approximation gives an
accurate result in a regime where the slow-roll approximation cannot
be used to compute the power spectrum.

\par

Another remark is in order. The fact that the WKB approximation is
better for large values of $\vert \beta \vert $ than for small values
of $\vert \beta +2\vert $ (scale invariance) might seem to be in
contradiction with Eqs.~(\ref{testdp}) and (\ref{testgw}), which
indicate that $\vert Q/\omega ^2\vert $ decreases as the slow-roll
parameters approach zero (scale invariance). In fact, there is no
contradiction because the WKB criterion only measures the accuracy
with which the mode function is predicted at a given time, without
taking into account the error coming from the matching procedure,
while the error in the power spectrum is a combination of these two
sources of error.

\par
 
\begin{figure}
\includegraphics[width=\linewidth]{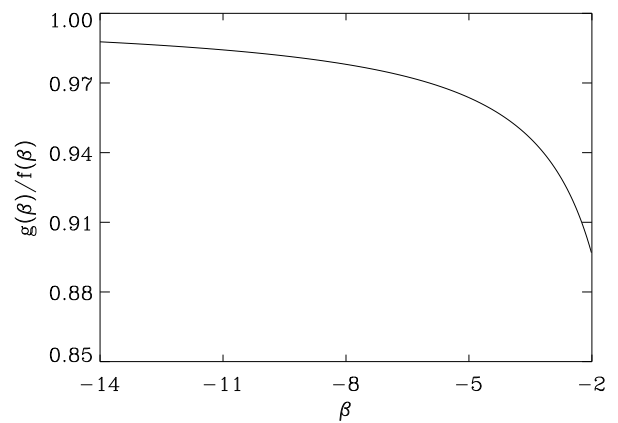}
\caption{Evolution of the ratio $g(\beta)/f(\beta)$ with
$\beta$. In the limit $\vert \beta \vert \rightarrow \infty $, the
ratio tends towards $1$. This can only be guessed from the figure because
the convergence rate is slow.}
\label{wkbamppl}
\end{figure}

The conclusion is that the WKB amplitude of the spectrum is always a
good approximation, see Fig.~\ref{wkbamppl}, although the precision
of the prediction for the amplitudes in the case of almost scale-invariant 
inflation is not as good as in the slow-roll approximation
\cite{SL,MS2,SG,LLMS}. Contrary to the latter, the spectral indices are
predicted exactly by the WKB approximation, $n_{_{\rm S}}-1 = n_{_{\rm
T}}=2\beta +4$. This is a big advantage if data sets of large
dynamic ranges are investigated.

\section{Application to slow-roll inflation}

Let us now compare the WKB approximation with the (leading-order)
slow-roll approximation. A comment is in order here on the meaning of
the phrases ``slow-roll conditions'' and ``slow-roll
approximation''. If the slow-roll conditions hold, all slow-roll
parameters are small. This does not specify the approximate scheme
that one should use in order to calculate the power spectra; in
particular it does not imply that the slow-roll approximation has to
be utilized in this situation. In this section, we will determine the
power spectra computed with the help of the WKB approximation in a
situation where the slow-roll conditions are valid.  Having clarified
this point, let us first quote the result of the slow-roll
approximation~\cite{SL,MS2}:
\begin{widetext}
\begin{eqnarray}
\label{srdp}
{\cal P}_{\zeta }(k) &=& \frac{H^2}{\pi\epsilon_1 m_{_{\rm Pl}}^2}
\left[1 - 2(C+1)\epsilon_1 - C\epsilon_2 
  - (2\epsilon_1 + \epsilon_2)\ln\left(\frac{k}{k_*}\right)
  + {\cal O}(\epsilon_n^2)\right]\, , 
\\
\label{srgw}
{\cal P}_h(k) &=& \frac{16 H^2}{\pi m_{_{\rm Pl}}^2}
\left[1 - 2(C+1)\epsilon_1 - 2\epsilon_1\ln\left(\frac{k}{k_*}\right) 
  + {\cal O}(\epsilon_n^2)\right]\, , 
\end{eqnarray} 
\end{widetext}
where $C\equiv \gamma _{_{\rm E}}+\ln 2-2\simeq - 0.7296$, $\gamma
_{_{\rm E}}\simeq 0.5772$ being the Euler constant. All quantities are
evaluated at $\eta_*$, which is the moment of time when $k_* =
aH(N_*)$.  The scale $k_*$ is called the pivot scale. It has been
shown in Ref.~\cite{MS2} that the corresponding time $\eta _*$ has a
nice interpretation: it is nothing but the time at which the matching
between the region where $aH\simeq k$ and the superhorizon region is
performed.  It is important to realize that, according to the previous
definition, $\eta _*$ is not a function of $k$, i.e. is the same
for all scales. Changing the time of matching is therefore equivalent
to changing the pivot scale. This means that the pivot scale must be
in the range of scales that are of astrophysical interest today but
otherwise can be chosen freely. It has been demonstrated in
Ref.~\cite{MS2} that the location of the pivot scale can be optimized
to reduce the slow-roll error in the power spectrum.

\par

{}From Eqs.~(\ref{srdp}) and (\ref{srgw}), we notice that close to the de
Sitter inflation the slow-roll approximation predicts the amplitude
for both density and tensor perturbations very well. The accuracy of
the slow-roll approximation improves as one approaches
scale-invariance, since this corresponds to $\epsilon_n \to 0$ (for
power-law inflation this corresponds to $\beta \to -2$). {\it A priori}, 
it gives an exact result for exact scale invariance but then the
linear approximation blows up, as can be seen from Eq.~(\ref{srdp}),
when $\epsilon \rightarrow 0$.

\par

In contrast, the WKB approximation does not give an exact result for
the amplitude in the scale-invariant limit (see the discussion of the
limit $\beta \to -2$ above), but it is able to give a reasonable
approximation in the regime where slow-roll is violated ($\epsilon_1
\to 1$). The spectral indices predicted by the slow-roll approximation
at first order are
\begin{equation}
\label{nsr}
n_{_{\rm S}}-1=-2\epsilon_1 -\epsilon_2 \, \quad n_{_{\rm T}}=-2\epsilon_1\, .
\end{equation}
These estimates are not exact, contrary to the WKB result in the case of 
power-law inflation, but provide good estimates close to the scale-invariant 
situation. 

\par

Let us now evaluate the WKB power spectra for the situation of slow-roll 
inflation. For that purpose we expand (\ref{wkbpspecdp}) and (\ref{wkbpspecgw})
at first order in the horizon flow functions and compare the result with 
(\ref{srdp}) and (\ref{srgw}). The effective frequencies read 
\begin{eqnarray}
\omega_{_{\rm S}}^2 &=& k^2\eta^2
   - \left(\frac{9}{4} + 3\epsilon_1 + \frac{3}{2}\epsilon_2\right)\, ,
\label{ws2} \\
\omega _{_{\rm T}}^2 &=& k^2\eta^2
   - \left(\frac{9}{4}+ 3\epsilon_1 \right)\, , 
\label{wt2}
\end{eqnarray}
and the WKB power spectra become 
\begin{eqnarray}
{\cal P}_\zeta &=& \frac{H^2}{\pi\epsilon_1 m_{_{\rm Pl}}^2}
(- k\eta_{\rm f})^3 \frac 23 \left(1 - \frac 83\epsilon_1 - 
\frac 13\epsilon_2\right)
{\rm e}^{2\Psi_{_{\rm S}}}, \\
{\cal P}_h &=& \frac{16 H^2}{\pi m_{_{\rm Pl}}^2}
(- k\eta_{\rm f})^3 \frac 23 \left(1 - \frac 83\epsilon_1\right)
{\rm e}^{2\Psi_{_{\rm T}}},
\end{eqnarray}
where we have made use of the relation $1/[(1-\epsilon_1)a H] = -
\eta[1 + {\cal O}(\epsilon_n^2)]$. The evaluation of the integrals in
the argument of the exponential functions gives
\begin{eqnarray}
& &\Psi (\eta_{\rm f}, \eta_*;k)= 
\int_{x_*}^{x_f} |\omega (y;k)| {\rm d}y \\
& &= - \int_{\eta_*(b)}^{\eta_{\rm f}}
\sqrt{\frac 94 - k^2\tau^2 + b(\epsilon,\delta,\xi)}{{\rm d}\tau\over\tau}\, ,
\end{eqnarray}
with
\begin{eqnarray}
\label{intpara}
b _{_{\rm S}}(\epsilon_n) &\equiv & 3\epsilon_1 
+ \frac{3}{2}\epsilon_2\, , 
\quad 
b _{_{\rm T}}(\epsilon_n) \equiv 3\epsilon_1 \, ,
\\
\eta_* (b) &=& - \frac{1}{k}\sqrt{\frac{9}{4} + b(\epsilon_n)}.
\end{eqnarray}
At first order, we may consider the horizon flow functions as being
constants. This is due to the fact that the derivatives of the
horizon flow functions are quadratic in these parameters. The
integration can be performed as for power-law inflation in the
previous section. The final result is
\begin{widetext}
\begin{eqnarray}
{\cal P}_\zeta (k)&=& \frac{H^2}{\pi\epsilon_1 m_{_{\rm Pl}}^2}
   (18 {\rm e}^{-3}) \left[1 - 2(D + 1)\epsilon_1 - D \epsilon_2
   - (2\epsilon_1 + \epsilon_2)\ln(-k\eta_{\rm f})
   + {\cal O}(\epsilon_n^2)\right]\, , \\
{\cal P}_h (k)&=& \frac{16 H^2}{\pi m_{_{\rm Pl}}^2}
   (18 {\rm e}^{-3}) \left[1 - 2(D + 1)\epsilon_1
   - 2\epsilon_1\ln(-k\eta_{\rm f}) + {\cal O}(\epsilon_n^2)\right]\, ,
\end{eqnarray}
with $D \equiv \frac 13 - \ln 3 \approx -0.765$, a reasonable
approximation of $C$ from above. All the quantities in the above
equation are evaluated at $\eta =\eta _{\rm f}$. In order to
compare this result with the slow-roll expressions (\ref{srdp}) and
(\ref{srgw}) we have to establish a link between $H(N_{\rm f})$ and
$H(N_*)$, $\epsilon_1(N_{\rm f})$ and $\epsilon_1(N_*)$, and so on. At
the level of a first-order expansion in the horizon flow
functions, we can consider $\epsilon_1$ and $\epsilon_2$ to be
constant everywhere, except in the factor $1/\epsilon_1$ in
(\ref{srdp}). Performing a linear expansion of $\epsilon_1$ around
$N_*$ we find
\begin{equation}
\frac{1}{\epsilon_1(N_{\rm f})} =  \frac{1}{\epsilon_1(N_*)} 
  [1 - \epsilon_2(N_*)\Delta N + \dots]. 
\end{equation}
In a similar manner we find 
\begin{equation}
H^2(N_{\rm f}) = H^2(N_*) [1 - 2\epsilon_1(N_*)\Delta N + \dots].
\end{equation} 
We eliminate $-\eta_{\rm f}$ in the logarithm by expressing it in
terms of $1/(aH)(N_{\rm f})$. Corrections containing
$\epsilon_1$ inside the logarithm can be omitted since they contribute
at higher order only. Finally, the logarithm can be written
\begin{equation}
\ln(- k \eta_{\rm f}) = \ln\left[{k\over (aH)(N_*)}\right] + 
                  \ln\left[{(aH)(N_*)\over (aH)(N_{\rm f})}\right] + \dots 
                = \ln\left({k\over k_*}\right)- \Delta N + \dots,
\end{equation} 
where now the dots denote terms of order $\epsilon_1$ and we have
used the pivot scale $k_* = (aH)(N_*)$.  Putting everything
together we find that all $\Delta N$ cancel, and the result simply
reads
\begin{eqnarray}
{\cal P}_\zeta (k)&=& \frac{H^2}{\pi\epsilon_1 m_{_{\rm Pl}}^2}
   (18 {\rm e}^{-3}) \biggl[1 - 2(D + 1)\epsilon_1 - D \epsilon_2
   - (2\epsilon_1 + \epsilon_2)\ln \biggl(\frac{k}{k_*}\biggr)
   + {\cal O}(\epsilon_n^2)\biggr]\, , \\
{\cal P}_h (k)&=& \frac{16 H^2}{\pi m_{_{\rm Pl}}^2}
   (18 {\rm e}^{-3}) \biggr[1 - 2(D + 1)\epsilon_1
   - 2\epsilon_1\ln \biggl(\frac{k}{k_* }\biggr) 
   + {\cal O}(\epsilon_n^2)\biggr]\, ,
\end{eqnarray}
\end{widetext}
which can now be compared with the prediction from the slow-roll power
spectra (\ref{srdp}) and (\ref{srgw}).

\par 

It is seen immediately that the spectral indices coincide with the
slow-roll result (\ref{nsr}). As already expected from the discussion
of power-law inflation above, the spectral indices are predicted by
the WKB approximation very well, the amplitudes only at an accuracy of
$10\%$ (coming from the factor $18{\rm e}^{-3}$). This is certainly
not good enough for future experiments, but it is well known that WKB
approximations can be systematically improved. 

\section{Conclusions}

A new method to predict inflationary power spectra has been presented,
based on the WKB approximation. One of the main advantages of this new
scheme of approximation is that it can give a good approximation even
if the slow-roll conditions are violated, whereas, up to now, all of
the methods in the literature were subject to this
limitation~\cite{remark}. We have checked this fact explicitly on the
example of power-law inflation. In addition, we have also tested the
method for the case of slow-roll inflation and find errors below
$10\%$ for all quantities considered. The prediction of the spectral
indices is exact in the case of the power-law inflation. Assuming that
the conclusions obtained from the power-law case are also valid in a
more general context, the WKB approximation appears as an efficient
method, especially for predicting the spectral indices. The main
restriction comes from the amplitude, which is not predicted with a
similar accuracy as the slow-roll approximation close to
scale invariance.

A possible disadvantage of the method is that it is non-local in the sense 
that an integration is involved from the time of horizon crossing
until some time when all the modes of interest are on superhorizon scales. 
This means, in contrast to the slow-roll approximation, where 
a certain behaviour of the background is assumed, that one needs to know
how the Hubble flow functions evolve with time in principle. In the example 
of slow-roll inflation it has been shown that one essential step in the WKB
approximation is to link the predictions at some time $\eta_{\rm f}$ to the
parameters relevant at the time of horizon crossing.
 
\par

Many improvements of the results presented in this article are
possible. As already mentioned above, the WKB expansion can be pushed
further, e.g. by going beyond the linear approximation for the joining 
of the super- and subhorizon WKB solutions. This will reduce the error in 
the amplitude, therefore
avoiding the main shortcoming of the new method. Another point is that
the standard slow-roll approximation (Bessel approximation) is valid
only when the slow-roll parameters are constant. Therefore, it cannot
be pushed beyond the first order. Recently, Stewart and Gong~\cite{SG}
have proposed a method, based on Green functions perturbative
calculations, which is free from this problem. They have calculated
the inflationary power spectra up to the second order. The WKB
approximation is also free from the above limitation and therefore
allows a determination of the spectra beyond the leading order. All
these questions are currently under investigation~\cite{MSwkb2}.

\vspace{0.3cm}

{\it Note added:} While the present article was being completed, a new
paper~\cite{H} appeared where the inflationary power spectra are also
calculated using a new method of approximation, namely the ``uniform
approximation'' according to the terminology used in that
article. Some of the equations obtained in~\cite{H} bear a close
resemblance to the results obtained in the present work as revealed by
a comparison of Eqs.~(\ref{wkbpspecdp}) and (\ref{wkbpspecgw}) with
Eq.~(14) of Ref.~\cite{H}. After having released the first version of
this paper, we became aware of an earlier attempt to use the WKB
approximation for cosmological perturbations by Nagasawa and Yokoyama
\cite{NY}.

\section*{Acknowledgements}
We thank F.~Finelli, M.~Nagasawa, C.~Terrero-Escalante and G.~Venturi for 
usefull comments on the first version and we acknowledge support from the
bilateral Austrian-French AMADEE program.


\begin{thebibliography}{50}
\bibitem{remark}{Exceptions are the constant and growing horizon 
approximations of Ref.~\cite{STG}. These approximations relax the slow-roll 
conditions by allowing that one of the horizon flow functions is much larger
than all the others, but still smaller than unity.}
\end{thebibliography}
\end{document}